\documentclass[%
  twocolumn,%
  prb,%
  showpacs,%
  letterpaper,%
  amsfonts,%
  floatfix,%
  balancelastpage%
]{revtex4}

\usepackage{amsmath,amssymb,amsbsy}
\usepackage{graphicx,subfigure}
\usepackage{hyperref}
\hypersetup{%
  breaklinks = {true},
  citecolor = {blue},
  colorlinks = {true},
  linkcolor = {red},
  pdfauthor = {\textcopyright\ Vitor M. Pereira},
  pdfcreator = {\LaTeX\ and \flqq hyperref\frqq},
  pdffitwindow = {true},
  pdfmenubar = {true},
  pdfpagelayout = {SinglePage},
  pdfstartview = {Fit},
  pdftoolbar = {true},
  plainpages = {false},
}

\newcommand{\bR}{\mathbf{R}}
\newcommand{\ba}{\mathbf{a}}
\newcommand{\bk}{\mathbf{k}}
\newcommand{\bq}{\mathbf{q}}
\newcommand{\Fig}{Fig.}
\newcommand{\Eq}{Eq.}
\newcommand{\Ref}{Ref.}
\newcommand{\tp}{t_\perp}
\newcommand{\vF}{\nu_F}
\newcommand{\ns}{n_e^*}
\newcommand{\nss}{n_e^{**}}

\begin{document}


\title{Distortion of the perfect lattice structure in bilayer graphene}


%

\author{Vitor M. Pereira}
\affiliation{Department of Physics, Boston University, 590
Commonwealth Avenue, Boston, MA 02215, USA\phantom{}}

\author{R. M. Ribeiro}
\affiliation{Center of Physics and Department of Physics,
University of Minho, P-4710-057, Braga, Portugal}

\author{N. M. R. Peres}
\affiliation{Center of Physics and Departament of Physics,
University of Minho, P-4710-057, Braga, Portugal}

\author{A.~H. Castro Neto}
\affiliation{Department of Physics, Boston University, 590
Commonwealth Avenue, Boston, MA 02215, USA}

\date{\today}


\begin{abstract}
We consider the instability of bilayer graphene with respect to a distorted
configuration in the same spirit as the model introduced by Su, Schrieffer and
Heeger. By computing the total energy of a distorted bilayer, we conclude that
the ground state of the system favors a finite distortion. We explore how the
equilibrium configuration changes with carrier density and an applied potential
difference between the two layers.
\end{abstract}

\maketitle


%
\section{Introduction} 
A planar arrangement of carbon atoms covalently bound via $sp^2$ orbitals
exhibits a honeycomb structure and is denoted graphene. Graphene rose rapidly
to the forefront of research in condensed matter physics, mostly because of the
peculiar electronic structure that emerges from its crystal structure, and the
consequent wealth of rich and unexpected phenomena\cite{CastroNeto:2007}.
Seminal experiments on two dimensional crystals established graphene as an
accessible reality\cite{Novoselov:2004}, and immediately unveiled numerous
surprises, both on a fundamental level -- like a new form of quantized Hall
effect -- and on a practical and technological level -- like the highly
efficient field effect and high electronic mobility
\cite{Novoselov:2005,Novoselov:2005b}. Most of the appealing phenomenology of
graphene owes to the fact that electron dynamics in this system can be described
in terms of chiral massless Dirac fermions\cite{Novoselov:2005} and, in fact,
graphene does exhibit many properties characteristic of relativistic
particles\cite{CastroNeto:2006b,Katsnelson:2007}.

Equally remarkable phenomena occur in bilayer graphene, which consists of two
adjacent graphene planes stacked in the A-B fashion typical of graphite.
Bilayer graphene displays the same sample quality and quasi-ballistic transport
characteristic of its single layer counterpart\cite{Morozov:2008}, but brings
also its share of new physics stemming from the nature of its charge
carriers: chiral massive electrons\cite{Novoselov:2006}. Most interesting is
the fact that, although gapless in its pristine form, a potential difference
between the two layers opens a gap in the spectrum that can be controlled via
chemical doping\cite{Ohta:2006} or
gating\cite{McCann:2006b,Castro:2007,Oostinga:2008}. 

Despite such favorable prospects, the amount of knowledge gathered in the
context of bilayer graphene still lags behind the intensity committed to single
layer graphene. In this brief paper, we address a particular
aspect of the electron-phonon interaction in bilayer graphene, namely the
tendency to relax the perfect crystal structure and generate a static, uniform
deformation. This effect is inspired, and similar in spirit, to the well known
Peierls instability that occurs in polyacetylene chains. As shown by Su,
Schrieffer and Heeger (SSH)\cite{Su:1979,Su:1980}, in polyacetylene the one
dimensional chain of carbon atoms has a half-filled electronic ground state that
is unstable with respect to a spontaneous dimerization. This dimerization opens
a sizeable gap in the spectrum that can be easily detected
experimentally\cite{Fincher:1978}.

\begin{figure}[b]
  \begin{center}
    \includegraphics*[width=0.46\textwidth]{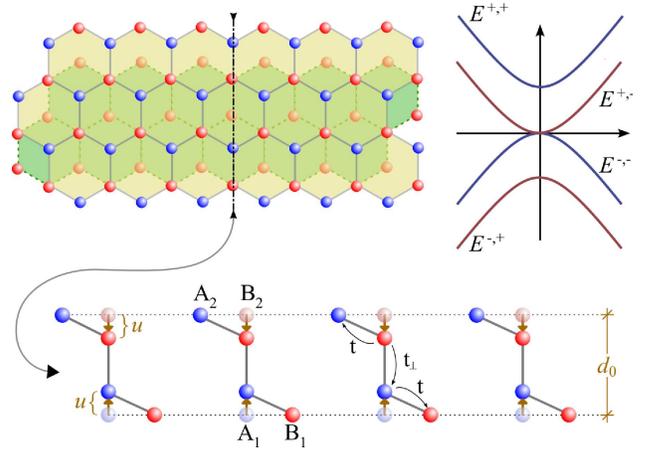}
  \end{center}
  \caption{(color online)
    Transverse view (bottom) along the dashed-dotted line (top left) 
    of the lattice distortion considered in the text.
    The vertical displacement of the carbon atoms connected by the
    hopping integral $t_\perp$ is represented by the parameter $u$. 
    In the top right we schematically represent the bandstructure 
    in the vicinity of the Dirac point.
  }
  \label{fig:Distortion}
\end{figure}
%

The instability we envisage for bilayer graphene is related with the
application of SSH's ansatz to the interplane hopping, $\tp$. 
From the outset, the atoms lying in the A and B sublattices within each layer
are not equivalent, since only one of the sublattices connects to the
adjacent plane (\Fig~\ref{fig:Distortion}). This has important experimental
consequences: one example is the known fact that in tunneling experiments one
typically detects only one of the sublattices of the topmost
layer\cite{Rutter:2007}.
In the absence of a potential difference between the two layers (unbiased
situation), the bilayer is a zero gap semiconductor, with hyperbolic bands
touching at the Fermi energy. A change in the interlayer hopping will not change
this situation, and thus the gap can still be tuned through the potential
difference between layers. However, as shown below, the electron-phonon
interaction might indeed lead to a stable distorted configuration.

%
\section{The model} 

\subsection{Tight binding description of a biased bilayer}

The electronic Hamiltonian of a biased bilayer
consists of two contributions, $H_e=H_{tb}+H_V$, where $H_{tb}$ is the
tight-binding Hamiltonian for the graphene bilayer, and $H_V$ reflects the
electrostatic bias applied between the two graphene planes.
The tight-binding Hamiltonian $H_{tb}$ contains in itself three terms describing
electron itinerancy among each individual plane and between the two
planes. In detail we have
\begin{equation}
  H_{tb} = H_{tb1}+H_{tb2}+H_\perp
  \,,
\end{equation}
with 
\begin{eqnarray}
  H_{tb1} =&-&t\sum_{\bR,\sigma}
  [
    a^\dag_{1\sigma}(\bR)b_{1\sigma}(\bR)
    +
    a^\dag_{1\sigma}(\bR)b_{1\sigma}(\bR-\ba_1)
    \nonumber\\
    &+&
    a^\dag_{1\sigma}(\bR)b_{1\sigma}(\bR-\ba_2)
    +
    \text{H.c.}
  ]\,,
\end{eqnarray}
\begin{eqnarray}
  H_{tb2} =&-&t\sum_{\bR,\sigma}
  [
    a^\dag_{2\sigma}(\bR)b_{2\sigma}(\bR)
    +
    a^\dag_{2\sigma}(\bR+\ba_1)b_{2\sigma}(\bR)
    \nonumber\\
    &+&
    a^\dag_{2\sigma}(\bR+\ba_2)b_{2\sigma}(\bR)
    +
    \text{H.c.}
  ]\,,
\end{eqnarray}
\begin{equation}
  H_\perp = -t_\perp\sum_{\bR,\sigma}
  [
    a^\dag_{1\sigma}(\bR)b_{2\sigma}(\bR)+
    b^\dag_{2\sigma}(\bR)a_{1\sigma}(\bR)
  ]\,,
\end{equation}
and
\begin{eqnarray}
  H_V =&& \frac V 2\sum_{\bR,\sigma}
  [
    a^\dag_{1\sigma}(\bR)a_{1\sigma}(\bR)+
    b^\dag_{1\sigma}(\bR)b_{1\sigma}(\bR)]
    \nonumber\\
    &-&\frac V 2\sum_{\bR,\sigma}
    [a^\dag_{2\sigma}(\bR)a_{2\sigma}(\bR)+
    b^\dag_{2\sigma}(\bR)b_{2\sigma}(\bR)
  ]
  \,.
\end{eqnarray}
In the above equations $\ba_1$ and $\ba_2$ represent the
elementary translations of the honeycomb lattice.
In the presence of an electrostatic bias $V$, the electronic dispersion
is given by the four branches
\begin{equation}
  E^{\pm,\pm}
  _{\bk}=\pm \frac 1 2 \sqrt{2t^2_\perp+V^2+4t^2\vert\phi_{\bk}\vert^2
  \pm \Delta_{\bk}
  }
  \,,
  \label{eq:dispV}
\end{equation}  
where 
\begin{equation}
  \Delta_{\bk}=
  2 \sqrt{t^4_\perp+4t^2(t^2_\perp+V^2)\vert\phi_{\bk}\vert^2}
  \,.
\end{equation}
When $V=0$ \Eq~\eqref{eq:dispV} simplifies to
\begin{equation}
  E^{\pm,\pm}
  _{\bk}=\pm \frac{1}{2}
  \Bigl(
    \pm\, t_\perp + \sqrt{t^2_\perp+4t^2\vert\phi_{\bk}\vert^2}
  \Bigr)
  \,,
  \label{eq:disp0}
\end{equation}
where $\phi_{\bk}$ is associated with the dispersion of a single layer, 
and is given by 
\begin{equation}
  \phi_{\bk} = 1 + e^{i\bk\cdot \ba_1}+e^{i\bk\cdot \ba_2}
  \,.
\end{equation}
In order to proceed analytically, we approximate $\vert \phi_{\bk}\vert$ by
\begin{equation}
  \vert \phi_{\bk}\vert \simeq \frac 3 2 a q
  \label{eq:DiracApprox}
  \,,
\end{equation}
where we took $\bk = \mathbf{K} + \bq$ to be close to the Dirac point in
the honeycomb Brillouin zone, and amounts to using the effective mass
approximation for bilayer graphene ($a$ is the carbon-carbon distance).

\begin{figure}[b]
  \centering
  \includegraphics[width=0.47\textwidth]{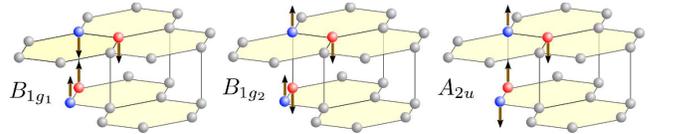}
  \caption{(color online)
    Phonon modes along the $c$-axis in graphite, using the notation of 
    \Ref~\onlinecite{Dresselhaus:2002}. The modes $A_{2u}$ and $B_{1g_2}$
    are nearly degenerate at the center of the Brillouin zone, with 
    $\omega\simeq 870\,\text{cm}^{-1}$ \cite{Dresselhaus:2002,Wirtz:2004}.
  }
  \label{fig:PhononModes}
\end{figure}
%

\subsection{Parametrization of Distortion}

We consider a distortion of the perfect lattice structure of the bilayer, such
that the $A$ and the $B$ atoms, connected by the hoping parameter $t_\perp$, 
distort along the vertical direction by an amount $u_i$
(\Fig~\ref{fig:Distortion}).  
In the spirit of the SSH model for polyacetylene\cite{Su:1980} we assume that,
to leading order in the strain, the effect of this
distortion is to change the value of $t_\perp$ according to
\begin{equation}
  t_\perp=t^0_\perp(1+\alpha u_i)
  \,,
  \label{eq:tp}
\end{equation}
where $t^0_\perp$ is the value of the interlayer hoping of
the undistorted lattice. For small $u_i$, the in-plane hopping $t$ is affected
by this distortion only at higher orders in $u_i$, and therefore we neglect its
variation.
This distortion will naturally induce an elastic
restoring force that we parametrize through the term 
\begin{equation}
  H_{el}=K\sum_{i=1}^{N_c}u^2_i+\sum_{i=1}^{N_c}\frac {P^2_i}{M}
  \,,
\end{equation}  
$N_c$ denoting the number of unit cells. In the static and homogeneous
situation the kinetic term gives an average null contribution and all $u_i$
acquire the same mean value: $u_i=u$. This leads to a total elastic energy that
reads:
\begin{equation}
  E_\text{el} = K\sum_{i=1}^{N_c}u^2 = N_c u^2
  \,.
  \label{eq:ElasticEnergy}
\end{equation}
The stability analysis of such a distorted phase
proceeds by
minimization of the total electronic and elastic energy, given by
$H_e+H_{el}$, with respect to the distortion $u$. We underline that, unlike 
in the original polyacetylene model\cite{Su:1980}, the parametrization
\eqref{eq:tp} does not change the original periodicity of the lattice, and
therefore does not require a density commensurability.

\subsection{Estimation of parameters}

A precise estimation of the parameters required for the computation of the
stable distorted configuration is not easy nor unique. On the one hand, little
is know with respect to the
structural and elastic properties of a graphene bilayer, and thus we will rely
on the corresponding knowledge that exists for graphite. On the other hand,
details like the type of substrate, can significantly alter these parameters, as
happens, for instance with the phonon spectrum that can be sensitive to
substrate and other constraints in the system. 

We will therefore resort to the structural parameters (lattice constant and
elastic coefficients) known for A-B stacked graphite. The carbon-carbon
distance is $a\simeq 1.42$~\AA, and the graphene unit cell has an area given by
$A_c=3\sqrt 3a^2/2$. The equilibrium interlayer distance is given by 
$d_0=c_0/2\simeq 3.35$~\AA, and corresponds to half the unit cell height of A-B
stacked graphite\cite{Jansen:1987,Boettger:1997,Mounet:2005}.

The value of the stiffness, $K$, can be estimated from the phonon spectrum of
graphite. In particular the $B_{1g_2}$ optical (out-of-plane) phonon mode has a
frequency of $\omega\approx 870\,\text{cm}^{-1}$, which is seen both
experimentally\cite{Nemanich:1977}, and from ab-initio calculations
\cite{Wirtz:2004}. As a result of the week interlayer interaction, this phonon
is essentially degenerate with the out-of-plane phonon $A_{2u}$ present in a
single layer of graphene. These normal modes are represented in
\Fig~\ref{fig:PhononModes}. 
We can assume that $K$ relates to this frequency through
$K a^2 \sim m \omega^2 a^2/ 4$, where $m$ is the carbon atom mass, and
$a\simeq 1.42$~\AA\ is the carbon-carbon distance\cite{Fuchs:2007}. As a result
we obtain as estimate for the stiffness $K\simeq8.5\,\text{eV.\AA}^{-2}$.

With respect to the electron-phonon coupling $\alpha$, its estimation is most
straightforward from the knowledge of how the interplane hopping varies with
distance. The interplane hopping, $\tp$ corresponds to the tight-binding
parameter $V_{pp\sigma}$ in the two center Slater-Koster 
formalism\cite{Slater:1954,Harrison:1999}. For instance, assuming that
\begin{equation}
  V_{pp\sigma}(r) \simeq A e^{-\alpha r}
  \label{eq:Slater-Koster}
\end{equation}
one can extract $\alpha$ from interpolation of the hoppings $\gamma_1$
the in-plane $V_{pp\sigma}(a)$ for graphite. Using the
values\cite{Dresselhaus:2002,Kim:2007} $\gamma_1\approx0.4~\text{eV}$ and
$V_{pp\sigma}(a)\approx 3.7~\text{eV}$ we obtain $\alpha \approx
1.2~\text{\AA}^{-1}$. Alternatively to the formula \eqref{eq:Slater-Koster}, one
could use a more refined interpolation formula for $V_{pp\sigma}(r)$ as
discussed in \Ref~\onlinecite{Papaconstantopoulos:1998}. This yields
$\alpha \approx
1.8~\text{\AA}^{-1}$, consistent with the previous estimate.

Finally, several recent experiments on the
bilayer\cite{Ohta:2007,Malard:2007,Yan:2007} show that the value of $\tp^0$ is
essentially the value expected in graphite, $\tp^0\approx 0.3~\text{eV}$, the
same applying to the in-plane hopping, $t\approx 3~\text{eV}$.

\begin{figure}[t]
  \begin{center}
    \includegraphics*[width=0.47\textwidth]{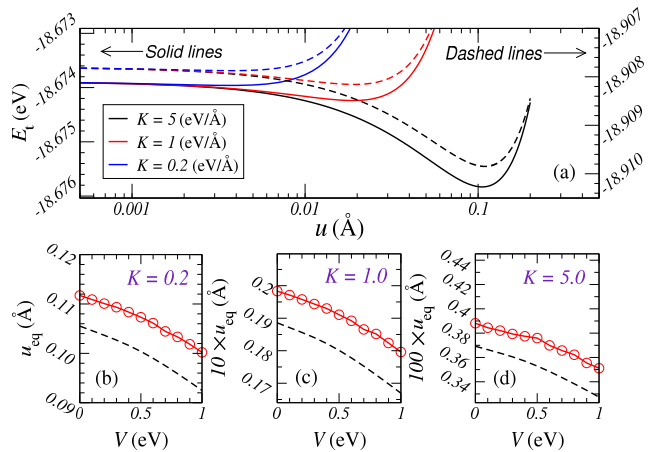}
  \end{center}
  \caption{(color online)
    (a) Total energy E$_t$ per unit cell as a function of the deformation
    parameter $u$, and for different values of $K$. The left (right) vertical 
    axis pertains to the solid (dashed) curves.
    (b-d) The equilibrium radius as a function of the bias voltage, $V$.
    In all panels dashed lines refer to the Dirac approximation, whereas full
    lines have been calculated using the full tight-binding dispersion in
    \Eq~\eqref{eq:dispV}. Notice that in panels (c) and (d) the vertical axis
    is amplified 10 and 100 times, respectively.
    Other parameters used are $t^0_\perp=0.3$ eV, $t=3.0$ eV, 
    $\alpha=1.5$\AA$^{-1}$.
  }
  \label{fig:ConstMu}
\end{figure}
%

%
\section{Ab-initio calculation of the elastic constant}
In addition to the above estimates of the model parameters, we have extracted
the compression elastic constant from a first principles calculation.
Density functional calculations in graphite and related compounds must be
carried with caution for it is known that different implementations of density
functional theory can yield noticeably different results
\cite{Wirtz:2004,Lazzeri:2008}.
Having this in mind, we calculated the equilibrium distance between graphene
planes in the bilayer by resorting to two different approximations: the
Generalized Gradient Approximation (GGA) and the Local Density Approximation
(LDA).

\emph{GGA ---} We sampled the BZ according to the scheme proposed by
Monkhorst-Pack \cite{Monkhorst:1968}, with a grid of $12\times 12\times
4$~$\mathbf{k}$-points. Bilayer graphene was modeled in a slab geometry by
including a vacuum region in a supercell containing 4 carbon atoms (2 for each
graphene sheet). 
In the normal direction ($z$-direction), the vacuum separating repeating slabs
has more than 10~\AA, and the size of the supercell in the $z$-direction was
optimized to make sure there was no interaction between repeating slabs. 

\emph{LDA ---} In this case the BZ was sampled within the same scheme with a
grid of $4\times 4\times 1$~$\mathbf{k}$-points, and using a supercell
comprising 8 carbon atoms (4 for each sheet). Adjacent slabs along the $z$
direction were separated by more than 30~\AA, and the size of the supercell
along this direction was again optimized.

In either case an increase in the number of sampling points did not result in a
significant total energy change, and the vertical separation quoted above
guarantees the absence of interaction between adjacent slabs. We used dual-space
separable pseudopotentials
by Hartwigsen, Goedecker, and Hutter\cite{Hartwigsen:1998} to describe the ion
cores. 
In a first step, all the atoms were fully relaxed to their equilibrium
positions. Then one of the graphene sheets was moved as a whole in the
$z$-direction by very small displacements, and the total energy of the system
was calculated, without any further relaxation.

\begin{figure}
  \centering
  \includegraphics*[width=0.47\textwidth]{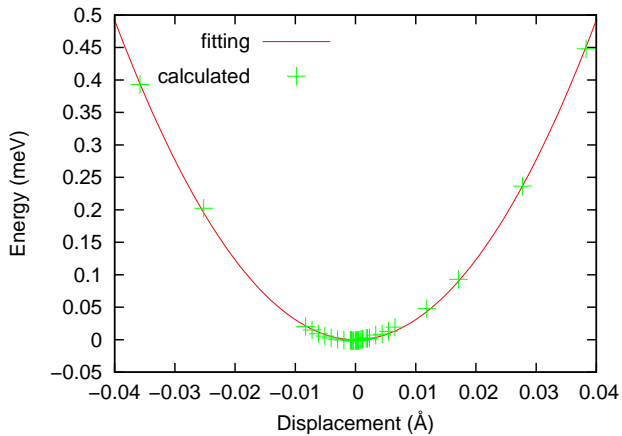}
  \caption{
    Energy as a function of distance between layers measured with respect to the
    equilibrium position, and calculated within the GGA.
    The positive direction indicates a displacement towards the other graphene
    sheet. The zero value energy is the energy of the fully relaxed sample.
  }
 \label{fig:enerXdist}
\end{figure}

Figure \ref{fig:enerXdist} shows the GGA variation of the total energy relative
to the relaxed sample, as a function of the displacement from the equilibrium
position. Also shown is the parabola that was fitted to the calculated values.
The fitting gave a value of $K=0.615\pm0.002$~eV/\AA$^2$, per unit cell.
The same calculation within LDA yields
$K=4.15\pm0.06$~eV/\AA$^2$. The two calculations therefore differ by one order
of magnitude, signaling the fact that, like other systems derived from
graphite, density functional calculations are very sensitive to the
details of the approximation used.

%
\section{Total energy at Constant $\mu$}
Our main objective is to quantify the equilibrium distortion that is expected
to emerge from the competition between elastic and electronic energies in the
ground state.
For illustration purposes we consider first the computation of the total energy
in the (potentially artificial) case where the chemical potential is held
constant. In particular, we assume that the density of carriers in the bilayer
is such that the chemical potential is located between $E^{+,+}_{q=0}$ and
$E^{+,-}_{q=0}$:
\begin{equation}
  \mu = \frac{V + \sqrt{V^2+4\tp^2}}{4}
  \,.
\end{equation}

Let us start with an unbiased bilayer ($V=0$). In this case the total electronic
energy per unit cell is given by
\begin{equation}
  \frac {E_e}{N_c} =
  \frac {g A_c}{2\pi} \int^{q_c}_0qdq \bigl(E_q^{-,+}+E_{q}^{-,-}\bigr)
  + \frac {g A_c}{2\pi} \int^{q_F}_0qdq E_q^{+,-}
  \,.
  \label{eq:TotEnConstMu}
\end{equation}
The integral is elementary leading to
\begin{align}
  \frac {E_e}{N_c} & = \frac{g A_c}{2\pi}\,
  \tp \biggl(
        \frac{q_t^2}{6} - \frac{q_F^2}{4} 
      \biggr)
  - \frac{g A_c}{6\pi}\,
  \tp q_t^2 \biggl[1+\biggl(\frac{q_c}{q_t}\biggr)^2 \biggr]^{3/2}\nonumber\\
  & + \frac{g A_c}{12\pi}\,
  \tp q_t^2  \biggl[1+\biggl(\frac{q_F}{q_t}\biggr)^2 \biggr]^{3/2}
  \,,
\end{align}
where the momenta $q_F$, $q_c$ and $q_t$ are defined as
\begin{equation}
  q_t = \frac{\tp}{3ta}\,,\;
  q_c = \sqrt{\frac{2\pi}{A_c}}\,,\;
  q_F = \frac{\sqrt{(2\mu+\tp)^2-\tp^2}}{3ta}\,,
\end{equation}
and $A_c=3\sqrt 3a^2/2$ is the area of the graphene unit cell. The total energy
$E_t$ per unit cell is given by
\begin{equation}
  \frac {E_t}{N_c}=\frac {E_e}{N_c} + Ku^2
  \,.
\end{equation} 
These two terms compete in such a way that the minimum energy state is
achieved for a finite value of $u$. 

The dashed lines of the top panels of \Fig~\ref{fig:ConstMu} represent the
total energy, $E_t$, as a function of the deformation $u$, using different
values of the stiffness parameter, $K$. It is also instructive to investigate to
what extent the approximation \eqref{eq:DiracApprox} influences the equilibrium
deformations and, for that, we have performed the calculation of the total
energies using the full tight-binding dispersions of \Eq~\eqref{eq:dispV}. The
results so obtained are represented in the same figure by the solid lines. 
It is clear from \Fig~\ref{fig:ConstMu}(a), that, besides yielding larger
slightly larger absolute values for the energy, the full dispersion increases
the equilibrium deformation by about 5 to 10\%.

The analytical calculation in the presence of a finite bias ($V\ne 0$), is also
straightforward. The total energy is still given by
\Eq~\eqref{eq:TotEnConstMu}, where $E_\bk^{\pm,\pm}$ is now given by
\eqref{eq:dispV}. The energy integrals are given in the appendix, the
final result being
\begin{equation}
  \frac{E_\text{el}}{N_c} = \frac{g A_c}{2\pi}
  \biggl[
    F^{-,+}(k)\Bigr|_0^{q_c} + F^{-,-}(k)\Bigr|_0^{q_c} +
    F^{+,-}(k)\Bigr|_0^{q_F}
  \biggr]
  \,.
\end{equation}
The primitives $F^{\eta_1,\eta_2}(k)$ are calculated in the appendix, with the
final result
\begin{align}
  F^{\eta_1,\eta_2}(k) &= \frac{\eta_1}{8\vF^2(\tp^2+V^2)} 
  \biggl\{
     \frac{R^{3/2}}{3\gamma} - \frac{\gamma x + \eta_2}{2\gamma^2}
      \eta_2 \sqrt{R} \nonumber\\
     &- \frac{\eta_2\Delta}{8\gamma^{5/2}} 
        \log\bigl(2\sqrt{\gamma R} + 2\gamma x + 2\eta_2 \bigr)
  \biggr\}
  \,,
\end{align}
and the remaining parameters are defined in \eqref{eq:EnIntParameters}.

Placing the chemical potential again at the midpoint
between the two conduction bands at $q=0$,
\begin{equation}
  \mu=(V+\sqrt{4t^2_\perp+V^2})/4
  \,,
\end{equation}
the corresponding Fermi wavevector is 
\begin{equation}
  q_F = \frac{1}{2\vF} 
  \sqrt{
    V^2+4 \mu + 2 \sqrt{
      4 \mu (\tp^2+V^2) - \tp^2 V^2
      }
    }
  \,,
\end{equation}
and we obtain the results shown in the lower 
panel of \Fig~\ref{fig:ConstMu} for the equilibrium radius. When $V$ varies
between 0 and 1~eV, the equilibrium radius shows a relative variation of $\sim
15\%$. In addition, it can be seen that the difference between using the Dirac
approximation and the full tight-binding dispersion is, in accordance with the
above, essentially a systematic increase in the equilibrium radius. For this
reason, henceforth we will restrict the discussion to the results obtained
within the Dirac approximation.

\begin{figure}
  \begin{center}
    \includegraphics*[width=0.46\textwidth]{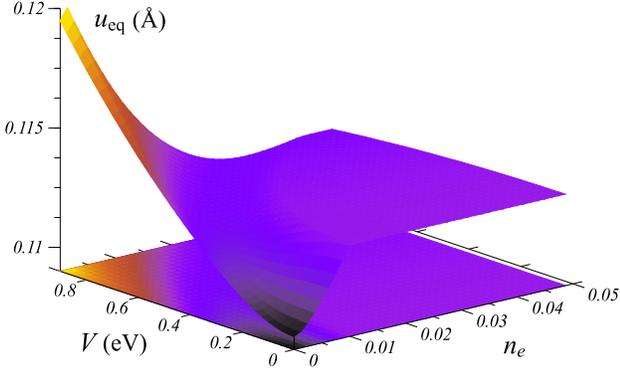}
  \end{center}
  \caption{(color online)
    Equilibrium deformation, $u_\text{eq}$, as a function of the electron
    density per unit cell ($n_e$) and the bias voltage ($V$). 
    The parameters used are $t^0_\perp=0.3~\text{eV}$, $t=3.0~\text{eV}$,
    $K=0.2~\text{eV.\AA}^{-2}$ and $\alpha=1.5~\text{\AA}^{-1}$.
  }
  \label{fig:UeqVsNeVsV}
\end{figure}
%

\begin{figure}[b]
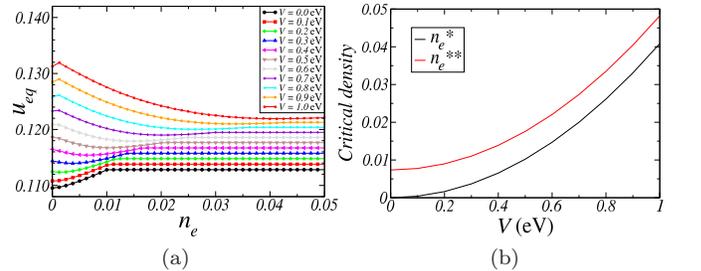

  \begin{center}
    \subfigure[][]{%
      \includegraphics*[width=0.5\columnwidth]{Figs/Ueq_vs_Ne}%
      \label{fig:UeqVsNe}%
    }%
    \subfigure[][]{%
      \includegraphics*[width=0.5\columnwidth]{Figs/ncStar}
      \label{fig:ncStar}%
    }
  \end{center}
  \caption{(color online)
    \subref{fig:UeqVsNe} 
    Selected cuts from \Fig~\ref{fig:UeqVsNeVsV} at constant bias $V$.
    For clarity, successive curves have been shifted vertically by 0.001 in the
    order of increasing $V$.
    \subref{fig:ncStar} 
    The densities $\ns$ and $\nss$ defined in   
    \Eq~\eqref{eq:CriticalDensities}, plotted as a function of $V$ for the 
    same parameters used in \Fig~\ref{fig:UeqVsNeVsV}.
  }
\end{figure}
%

%
\bigskip
\section{Total Energy at Constant $\mathbf{n_e}$}
We consider now the more relevant case of a bilayer with constant carrier
density, which can be tuned, for instance, through a gate voltage. We define
$n_e$ as the number of electrons per unit cell, with respect
to the charge-neutral situation in which the valence bands are fully occupied.
In addition we will be concerned with electron doping only. The calculation of
the electronic energy in this case requires, in general, the consideration of
three distinct possibilities. Assuming a biased situation, and with
respect to the notation defined in \Fig~\ref{fig:BandParameters}, we can have:
\begin{enumerate} \renewcommand{\labelenumi}{(\roman{enumi})}
  \item the Fermi level lying between $E_1$ and $E_2$, in which case the Fermi
        surface consists of a Fermi ring characterized by two Fermi momenta
        $q_F^{III,1}$ and $q_F^{III,2}$, and the phase space exhibits a central
        hollow;
  \item the Fermi level lying between $E_2$ and $E_3$, where we have a more
        conventional Fermi surface;
  \item the Fermi level lying above the bottom of the uppermost band, in which
case
        we have again two Fermi momenta, $q_F^{III}$ and $q_F^{IV}$, but the
        phase space is now simply connected.
\end{enumerate}
The boundaries of these regimes can be easily identified through the two
threshold densities 
\begin{equation}
  \ns = \frac{g A_c}{4\pi}\, q_2^2
  \,,\quad\text{and}\quad
  \nss = \frac{g A_c}{4\pi}\, q_3^2
  \label{eq:CriticalDensities}
  \,.
\end{equation}
It follows that the total electronic energy is computed as
\begin{align}
  \frac {E_e}{N_c} & =
  \frac {g A_c}{2\pi} \int^{q_c}_0qdq \bigl(E_q^{-,+}+E_{q}^{-,-}\bigr) \\
  & + \frac {g A_c}{2\pi} \int^{q_F^{III,2}}_{q_F^{III,1}}qdq E_q^{+,-}
  + \frac {g A_c}{2\pi} \int^{q_F^{IV}}_{0}qdq E_q^{+,+} \nonumber
  \,,
\end{align}
where the integration limits of the last two terms are given by
(see also \Fig~\ref{fig:BandParameters} for notation)
\begin{subequations}
\begin{widetext}
  \begin{align}
    n_e<\ns: & \\ 
    & q_F^{IV} = 0 \,,\nonumber\\
    & 
    \begin{cases}
      n_e = \frac{g A_c}{4\pi} \biggl(
      \Bigl(q_F^{III,2}\Bigr)^2 - \Bigl(q_F^{III,1}\Bigr)^2 \biggr) \\
      2 \vF^2 \Bigl(q_F^{III,1}\Bigr)^2 - \sqrt{\tp^4 + 4 \vF^2
      \Bigl(q_F^{III,1}\Bigr)^2 \bigl(\tp^2 + V^2\bigr)} = 
      2 \vF^2 \Bigl(q_F^{III,2}\Bigr)^2 - \sqrt{\tp^4 + 4 \vF^2
      \Bigl(q_F^{III,2}\Bigr)^2 \bigl(\tp^2 + V^2\bigr)}
    \end{cases} \nonumber \\
    \ns<n_e<\nss: & \\ 
    & q_F^{IV} = 0 \,,\nonumber\\
    & q_F^{III,1} = 0 \,,\nonumber\\
    & q_F^{III,2} = \frac{g A_c}{4\pi} \biggl(\frac{1}{n_e}\biggr)^{-1/2}
      \,.\nonumber\\
    n_e>\nss: & \\ 
    & q_F^{III,1} = 0 \,,\nonumber\\
    & 
    \begin{cases}
      n_e = \frac{g A_c}{4\pi} \biggl(
        \Bigl(q_F^{III,2}\Bigr)^2 + \Bigl(q_F^{IV}\Bigr)^2 
      \biggr) \nonumber \\
      2 \vF^2 \Bigl(q_F^{III,2}\Bigr)^2 + \sqrt{\tp^4 + 4 \vF^2
      \Bigl(q_F^{III,2}\Bigr)^2 \bigl(\tp^2 + V^2\bigr)} = 
      2 \vF^2 \Bigl(q_F^{IV}\Bigr)^2 - \sqrt{\tp^4 + 4 \vF^2
      \Bigl(q_F^{IV}\Bigr)^2 \bigl(\tp^2 + V^2\bigr)}
    \end{cases} \nonumber 
  \end{align}
\end{widetext}
\end{subequations}

Minimizing the total energy with respect to $u$ yields the equilibrium
displacements plotted in \Fig~\ref{fig:UeqVsNeVsV}, for different electron
densities and bias voltages. The typical deformations for the parameters
quoted in the figure are $\sim 0.11~\text{\AA}$, which
represents $\sim 8$\% of the carbon-carbon distance, $a$. The variation of
$u_\text{eq}$ with $n_e$ and $V$ is non-monotonic. In particular, one notices
that, for constant $V$, the equilibrium deformation tends to saturate beyond a
given density. This can be appreciated in more detail in
\Fig~\ref{fig:UeqVsNe}, where we present selected cuts of the same surface. The
saturation can be understood from the interplay of two factors: on the one hand,
the variation of $V$ and $n_e$ induce changes in the bandstructure only in a
region close to the neutrality point; on the other hand, for high enough
density, the Fermi level will always be considerably above the bottom of the
uppermost band ($E_3$ in \Fig~\ref{fig:BandParameters}). In fact, comparing the
values of $\ns$ and $\nss$ presented in \Fig~\ref{fig:ncStar}, one can verify
that the first sets the scale for the minimum in the curves of $u_\text{eq}$
versus $n_s$, therefore  defining the shape of the valley in the plot of
\Fig~\ref{fig:UeqVsNeVsV}. The value $\nss$, on the other hand, marks the
onset of saturation.

%
\section{Discussion and Conclusions}
We have shown that a graphene bilayer with A-B stacking can be unstable
with respect to a Peierls-like distortion affecting the interplane bonds. This
distortion preserves the bandstructure of the system, in the sense that,
unlike the original Peierls problem, it does 
not lead to a gap in the unbiased case, nor to its closing in the biased
situation. In addition, it was found that the general effect of the bias
voltage is to increase the equilibrium deformation.

By comparing the results obtained with the full
tight-binding dispersion of bilayer graphene with the effective mass
approximation [\Fig~\ref{fig:ConstMu}(c-e)], we concluded that the
former does not introduce significant changes in the equilibrium results, and
therefore the low energy approximation is adequate to study this instability.

For the values of $K$ used in \Fig~\ref{fig:UeqVsNeVsV}, the
magnitude of the deformation corresponds to roughly $10$\% of the in-plane
carbon-carbon distance, and is significant. However, at this point one can
hardly be definite about a specific value of the equilibrium deformation on
account of the uncertainties in the estimation of the parameters $K$ and
$\alpha$. 
The value used for $K$ is close to the compressive stiffness found with the GGA
calculation described above. But clearly, had we used the estimate for the
phonon $B_{1g_2}$ (or the LDA result) instead, we would have obtained much
smaller values of $u_\text{eq}$, as can be inferred
from \Fig~\ref{fig:ConstMu}(d), although the qualitative 
features of \Fig~\ref{fig:UeqVsNeVsV} would be preserved. 
Hence a definitive conclusion as to the magnitude of the effect is deferred
until the relevant parameters in bilayer graphene are experimentally available.

In the consideration of the electronic energy, he have accounted only for
nearest neighbor in-plane and interplane hoppings. Additional hopping terms,
like next-nearest neighbor and other interplane hoppings, should not change the
qualitative picture presented here. On a quantitative level, even the ones
that are affected in first order in $u$ are expected to contribute only slightly
on account of their smaller magnitudes in comparison with $t$ and $\tp$.

%
%
\acknowledgments
V.M.P. is supported by Funda\c{c}\~ao para a Ci\^encia e a
Tecnologia (FCT) via SFRH/BPD/27182/2006, and acknowledges Centro
de F{\'i}sica do Porto for computational support. N.M.R.P and V.M.P
acknowledge the support of POCI 2010 via PTDC/FIS/64404/2006.
R.M.R. acknowledges the support of FCT under the SeARCH (Services and Advanced
Research Computing with HTC/HPC clusters) project (contract CONC-REEQ/443/2005).

%
%
\appendix

%
\section{Bandstructure Parameters}

\begin{figure}
  \begin{center}
    \includegraphics[width=0.8\columnwidth]{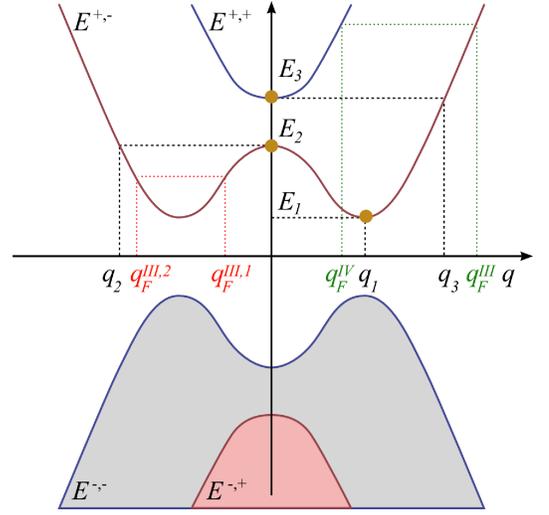}
  \end{center}
  \caption{(color online)
    Schematic representation of an arbitrary cut of the bandstructure of bilayer
    graphene close to the Dirac point.
  }
  \label{fig:BandParameters}
\end{figure}

With respect to the bandstructure depicted in \Fig~\ref{fig:BandParameters},
the notable momenta are ($\vF = 3ta/2$):
\begin{subequations}\label{eq:NotableMomenta}
\begin{align}
  q_1 &= \frac{1}{2\vF}\sqrt{\frac{V^4 + 2\,V^2 \tp^2}{\tp^2+V^2}}\,, \\
  q_2 &= \frac{V}{\vF}\,, \\
  q_3 &= \frac{1}{\vF} \sqrt{V^2+2\, \tp^2}\,,
\end{align}
\end{subequations}
while the corresponding energies are
\begin{subequations}\label{eq:NotableEnergies}
\begin{align}
  E_1 &= \frac{1}{2}\frac{\tp V}{\sqrt{V^2+\tp^2}}\,, \\
  E_2 &= \frac{V}{2}\,, \\
  E_3 &= \frac{1}{2} \sqrt{V^2+4\, \tp^2}\,.
\end{align}
\end{subequations}
The energy gap is given by 
\begin{equation}
  \Delta = 2 E_1 = \frac{\tp V}{\sqrt{V^2+\tp^2}}
  \,,
\end{equation}
and the midpoint between the upper bands at $q=0$ is at 
\begin{equation}
  E_\text{mid} = \frac{V + \sqrt{V^2 + 4\,\tp^2}}{4}
  \,,
\end{equation}
to which corresponds the momentum
\begin{equation}
  q_{E_\text{mid}} = \frac{1}{2\vF} 
  \sqrt{
    V^2+4 E_\text{mid} + 2 \sqrt{
      4 E_\text{mid} (\tp^2+V^2) - \tp^2 V^2
      }
    }
  \,.
\end{equation}
%

%
%

%
\section{Energy Integrals}
To compute the total electronic energy, the evaluation of the integral 
\begin{equation}
  F^{\pm,\pm}\equiv\pm\frac{1}{2}\int k dk \sqrt{A + B k^2 \pm 2 \sqrt{D+Ek^2}}
\end{equation}
is required. The parameters, with respect to the dispersion of the bilayer
in \Eq~\eqref{eq:dispV}, are given as
\begin{gather}
  A = V^2 + 2 \tp^2 \,,\quad
  B = 4\vF^2 \,,\quad
  D = \tp^4  \,,\nonumber\\
  E = 4\vF^2(tp^2+V^2) \,.
\end{gather}
The integral is readily computed by changing to the variable 
$x\equiv\sqrt{D + E k^2}$, after which it becomes
\begin{equation}
  \int xdx \sqrt{\alpha + \beta x + \gamma x^2}
  \,,
\end{equation}
and is readily available in standard tables. The final result is thus
\begin{align}
  F^{\eta_1,\eta_2} &= \frac{\eta_1}{4\vF^2(\tp^2+V^2)} 
  \biggl\{
     \frac{R^{3/2}}{3\gamma} - \frac{\gamma x + \eta_2}{2\gamma^2}
      \eta_2 \sqrt{R} \nonumber\\
     &- \frac{\eta_2\Delta}{8\gamma^{5/2}} 
        \log\Bigl(2\sqrt{\gamma R} + 2\gamma x + 2\eta_2 \Bigr)
  \biggr\}
  \,,
\end{align}
where
\begin{align}
  R & \equiv \alpha + \beta x + \gamma x^2 \,,\nonumber\\
  \Delta & \equiv 4\alpha\gamma - \beta^2 = \frac{4V^2\tp^2}{(\tp^2+V^2)^2}
    \,,\nonumber\\
  \alpha & \equiv A-\frac{BD}{E} = \frac{V^4+\tp^4+3\tp^2V^2}{\tp^2+V^2}
    \,,\nonumber\\
  x &\equiv\sqrt{\tp^4 + 4\vF^2(\tp^2+V^2) k^2}\,,\nonumber\\
  \gamma &= \frac{1}{\tp^2+V^2} \,,\nonumber\\
  \beta &= \pm 2
  \,.
  \label{eq:EnIntParameters}
\end{align}
%


\bibliographystyle{apsrev}
\bibliography{bilayer-peierls}

\end{document}